\documentclass[twocolumn,prl,amsmath,letterpaper,floatfix,showpacs]{revtex4}
\usepackage{graphicx}
\usepackage{amsmath}
\usepackage{amssymb}

\begin{document}

\title{Noise auto-correlation spectroscopy with coherent Raman scattering}
\date{\today}

\author{Xiaoji G. Xu, Stanislav O. Konorov, John W. Hepburn \& Valery Milner}

\affiliation{Department of Chemistry and the Laboratory for Advanced Spectroscopy and Imaging
Research (LASIR), University of British Columbia, 2036 Main Mall, Vancouver, BC, Canada V6T 1Z1.}

\begin{abstract}
Ultrafast lasers have become one of the most powerful tools in coherent nonlinear optical spectroscopy. Short pulses enable direct observation of fast molecular dynamics\cite{Zewail94, Schmitt97}, whereas broad spectral bandwidth offers ways of controlling nonlinear optical processes\cite{Meshulach98,Dudovich02} by means of quantum interferences\cite{Shapiro03}. Special care is usually taken to preserve the coherence of laser pulses as it determines the accuracy of a spectroscopic measurement. Here we present a new approach to coherent Raman spectroscopy based on deliberately introduced noise, which increases the spectral resolution, robustness and efficiency. We probe laser induced molecular vibrations using a broadband laser pulse with intentionally randomized amplitude and phase. The vibrational resonances result in and are identified through the appearance of intensity correlations in the noisy spectrum of coherently scattered photons. Spectral resolution is neither limited by the pulse bandwidth, nor sensitive to the quality of the temporal and spectral profile of the pulses. This is particularly attractive for the applications in microscopy\cite{Cheng04}, biological imaging\cite{Cheng02} and remote sensing\cite{Ooi05}, where dispersion and scattering properties of the medium often undermine the applicability of ultrafast lasers. The proposed method combines the efficiency and resolution of a coherent process with the robustness of incoherent light. As we demonstrate here, it can be implemented by simply destroying the coherence of a laser pulse, and without any elaborate temporal scanning or spectral shaping commonly required by the frequency-resolved spectroscopic methods with ultrashort pulses.
\end{abstract}

\maketitle

\begin{figure}
\includegraphics[width=1.0\columnwidth]{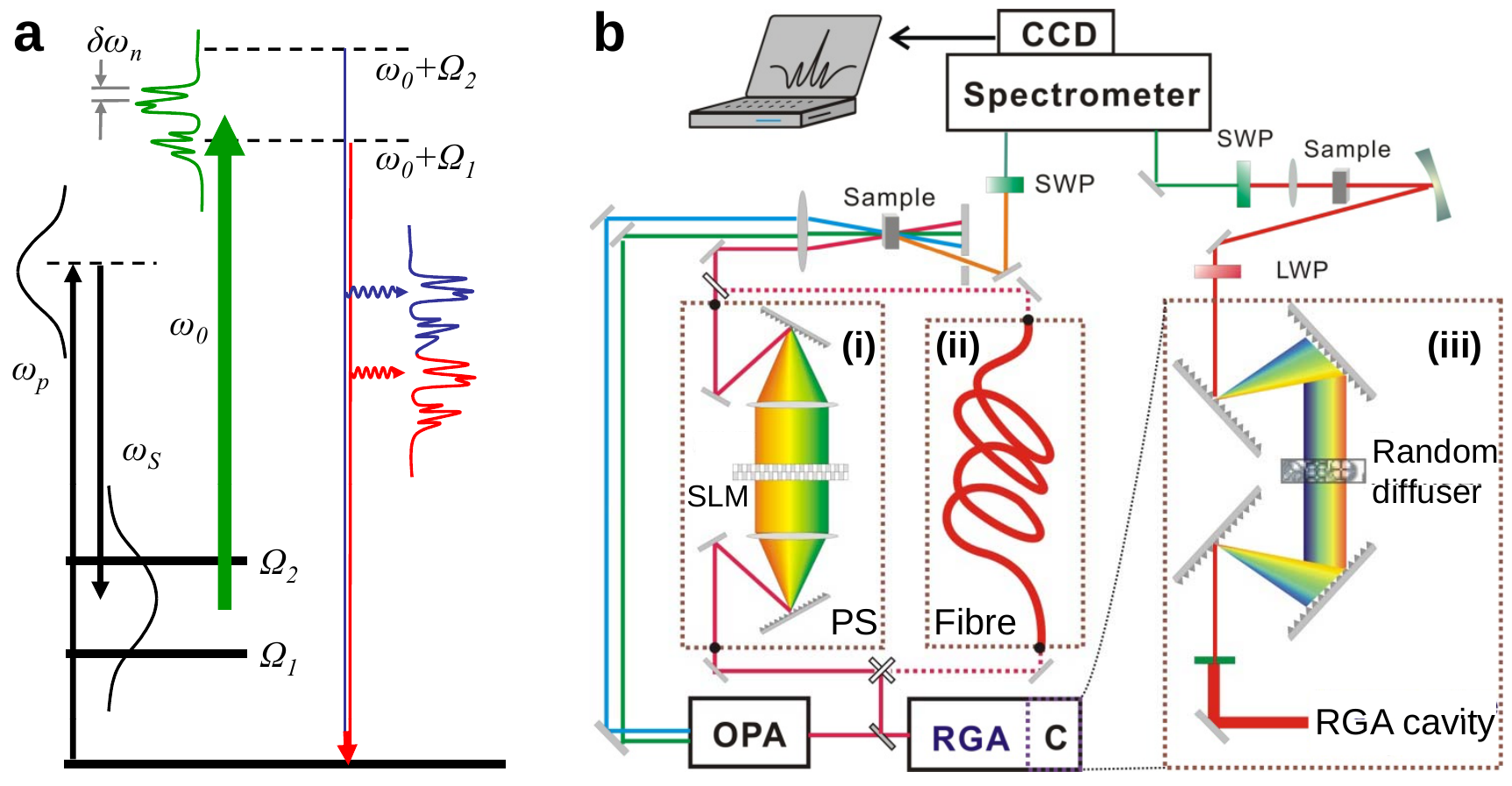}
\caption{
Interaction scheme for coherent anti-Stokes Raman scattering (CARS) and experimental arrangement for the detection of CARS signal. \textbf{a}, Ultra-short pump and Stokes pulses (black) overlap in time and prepare the coherent molecular vibrations covered by the collective pump-Stokes frequency bandwidth. The broadband probe pulse (green) interacting with the vibrational energy levels $\Omega _{1,2}$ generates the anti-Stokes light (red and blue). In the method of noise auto-correlation spectroscopy with coherent anti-Stokes Raman scattering (NASCARS), the noise in the probe spectrum is imprinted onto the corresponding spectral regions of the anti-Stokes light. Correlations in the anti-Stokes spectrum (red-blue curve on the right) reveals the vibrational beating frequencies with the resolution dictated by the noise correlation length $\delta \omega _{n}$. \textbf{b}, Diagrams of various experimental configurations. In the multi-color schemes \textbf{(i)} and \textbf{(ii)}, the probe pulses are generated by a regenerative Ti:Sapphire amplifier (RGA), while the pump and Stokes pulses are produced by an optical parametric amplifier (OPA). The pulses are focused on the sample in the non-collinear \textsc{boxcars} geometry. After filtering the anti-Stokes pulse with a short-wave pass filter (SWP), its spectrum is detected by the CCD-based spectrometer, and the auto-correlation of the spectrum is calculated by the computer. In (i), the white spectral noise is applied to the probe pulse by means of the pulse shaper (PS). In (ii), the phase of the probe is randomized due to the inter-modal dispersion of a multi-mode optical fibre. In the single-pulse configuration, a diffuser is introduced in the compressor stage (C) of the laser system, schematically depicted in \textbf{(iii)}(actual layout depends on the design of the compressor and is usually implemented with a single diffraction grating). Random scattering results in the partial de-coherence of the pulse required by NASCARS. After passing a long-wave pass filter (LWP), the beam is focused on the sample with a concave mirror. Different realizations of noise are executed either by re-programming the spatial light modulator (SLM) or simply by moving the fibre or the diffuser between each data acquisition.}
\label{Fig1}
\end{figure}

We apply our method to coherent anti-Stokes Raman scattering (CARS), which has become the method of choice in nonlinear optical spectroscopy and microscopy\cite{Zheltikov00, Cheng04} after the advent of powerful ultrafast lasers. As a third-order nonlinear process, femtosecond CARS exhibits high efficiency at low average laser power. High sensitivity to molecular structure enables detection of small quantities of complex molecules\cite{Scully02} and non-invasive biological imaging\cite{Cheng02}. In CARS, pump and Stokes photons of frequencies $\omega _{p}$ and $\omega _{S}$, respectively, excite molecular vibrations at frequency $\Omega = \omega _{p}-\omega _{S}$ (Fig.~\ref{Fig1}a). A probe photon at $\omega _{0}$ is scattered off the coherent vibrations generating the anti-Stokes signal at frequency $\omega=(\omega _{0} + \Omega)$. Temporal and spectral resolution of CARS spectroscopy is typically limited by the duration of the excitation pulses and their frequency bandwidth, respectively. Broadband femtosecond pulses, though advantageous for time-resolved CARS spectroscopy\cite{Schmitt97,Lang99}, offer poor spectral resolution. The latter can be improved by means of the pulse shaping technique\cite{Oron02,Xu07} which modifies the amplitudes and phases of the spectral constituents of the pulse\cite{Weiner00} (Fig.~\ref{Fig1}b(i)). It enables selective excitation\cite{Dudovich02,Pestov07} or selective probing\cite{Oron02a,Lim05} of separate vibrational modes on the frequency scale narrower than the overall pulse bandwidth. Selective excitation is based on the coherent control of nonlinear optical processes, while selective probing employs the idea of multiplex CARS in which a small part of the pulse spectrum is modified (in amplitude, phase or polarization) and serves as a narrowband probe. Both approaches rely on the effects of quantum or optical interference, and are therefore sensitive to noise and de-coherence which often ruin the spectral and temporal integrity of the pulse. Using a narrow slice of the available pulse spectrum as a probe in multiplex CARS also means that higher resolution may only be obtained at the expense of the proportionally lower probe power and correspondingly lower signal.

In this work we show that contrary to the common belief that spectral noise is detrimental to coherent spectroscopic measurements, it can be utilized for improving the resolution, efficiency and robustness against unavoidable degradation of pulse coherence. In our new method of noise auto-correlation spectroscopy with coherent anti-Stokes Raman scattering (NASCARS) the laser-induced molecular vibrations are probed by a broadband pulse with intentionally randomized amplitude and phase. Due to the presence of Raman resonances, the spectral noise of the probe field is transferred to the spectrum of the scattered photons. While completely uncorrelated in the probe, the noise in the anti-Stokes field acquires strong correlations as a result of the molecular vibrational coherence. The latter assures that multiple Raman modes ``map'' the same realization of the probe noise onto the different spectral regions of the generated anti-Stokes spectrum, as shown schematically in Fig.~\ref{Fig1}a. An auto-correlation of the noisy anti-Stokes spectrum corresponds directly to the auto-correlation of the Raman mode structure, thus revealing the vibrational beating frequencies of the medium. Even though the probe in NASCARS carries the full available spectral bandwidth of a femtosecond pulse, the resolution is determined by the much smaller frequency correlation length of the introduced noise,\ $\delta \omega _{n}$. Utilization of the full bandwidth provides the ability to increase the resolution with no penalty in the available probe power and, therefore, with no direct reduction of the signal level.

Noise and fluctuations in resonant fluorescence\cite{Lu97}, scattered light\cite{Schrof98} and number of trapped ultra-cold atoms\cite{Rom06} have been used to observe statistical properties of a medium but did not provide any spectroscopic information. NASCARS extends the idea of using noise correlations as a diagnostic tool to the domain of coherent nonlinear spectroscopy with ultrashort pulses. Noisy nanosecond lasers have been previously exploited in coherent Raman spectroscopy as a means of achieving time resolution shorter than the duration of the excitation pulses\cite{Stimson96,Stimson97}. The method of coherence observation by interference noise (COIN\cite{Kinrot95}) has been introduced for detecting the decay rates of electronic coherence, and applied to linear spectroscopy with narrowband picosecond pulses and to quantum state holography\cite{Averbukh99}. Unlike NASCARS, both approaches retrieve spectroscopic data from scanning the time delay between the noisy pulses. Here we show that the availability of a broad spectrum advances the method of noise-assisted spectroscopy towards efficient scan-less single-beam frequency-resolved analysis of molecular vibrations.

Consider the scheme depicted in Fig.~\ref{Fig1}a with two Raman modes at frequencies $\Omega _{1}$ and $\Omega _{2}$ excited by the pump-Stokes pulse pair.  The anti-Stokes field $E(t)$, generated by the Raman scattering of the probe field $E_{0}(t)$, can be expressed as the real part of $E(t)=E_{0}(t)\cdot e^{-\gamma t}\left[ e^{-i\Omega_{1}t} + e^{-i\Omega_{2}t} \right]$, where for the sake of clarity we have neglected the non-resonant background, and assumed equal strength and decay rate $\gamma $ for both vibrational modes. The experimentally observed anti-Stokes spectrum $I(\omega )$ is the Fourier transform of the field correlation function $G_{E}(\tau )$, where
\begin{equation}\label{GsubEoftau}
    G_{E}(\tau )=\int{dtE(t)E^{*}(t-\tau )} = G_{\mathcal{E}_{0}}(\tau )[e^{i\Omega_{1}\tau} + e^{i\Omega_{2}\tau}].
\end{equation}
Here $G_{\mathcal{E}_{0}}(\tau )$ is the correlation function of the modified input probe field $\mathcal{E}_{0} \equiv E_{0}\cdot e^{-\gamma t}$, i.e. the probe field multiplied by the exponential decay of the vibrational coherence. Hence,
\begin{eqnarray}\label{Iofomega}
I(\omega ) &=& \int{d\tau e^{-i\omega \tau }G_{\mathcal{E}_{0}}(\tau )[e^{i\Omega_{1}\tau} + e^{i\Omega_{2}\tau}]}
\nonumber\\
&=& \mathcal{I}_{0}(\omega -\Omega_{1})+\mathcal{I}_{0}(\omega -\Omega_{2}),
\end{eqnarray}
where $\mathcal{I}_{0}(\omega -\Omega_{1,2})$ is the spectrum of $\mathcal{E}_{0}$. Equation (\ref{Iofomega}) shows that any amplitude noise in $\mathcal{I}_{0}$ is mapped onto the corresponding spectral region of the anti-Stokes spectrum $I$ (that is, up-shifted by the corresponding Raman shift $\Omega _{1,2}$ as schematically shown in Fig. \ref{Fig1}a). Averaging over multiple noise realizations, we arrive at the expression for the noise auto-correlation for the measured anti-Stokes spectrum as a function of the frequency shift $\Delta \omega$:
\begin{equation}\label{GsubIaverage}
    \langle G_{I}(\Delta \omega )\rangle=G_{\mathcal{I}_{0}}\cdot [2g(\Delta \omega ) + g(\Delta \omega +\delta \Omega)+ g(\Delta \omega -\delta \Omega)],
\end{equation}
where $\delta \Omega \equiv \Omega_{2}-\Omega_{1}$ is the distance between the vibrational resonances, and we have assumed an uncorrelated probe noise with $G_{\mathcal{I}_{0}}(\Delta \omega )\equiv G_{\mathcal{I}_{0}} g(\Delta \omega )$ and $g(\Delta \omega )$ being the line shape function peaked at $\Delta \omega =0$ and of width $\delta \omega _{n}$. The two sidebands at $\pm \delta\Omega $ in Eq.~\ref{GsubIaverage} illustrate the ability of NASCARS to retrieve the vibrational beating frequency from the frequency correlations in the noisy spectrum of the Raman-scattered light, with the resolution determined by the noise correlation length $\delta \omega _{n}$.

\begin{figure}
\includegraphics[width=1.0\columnwidth]{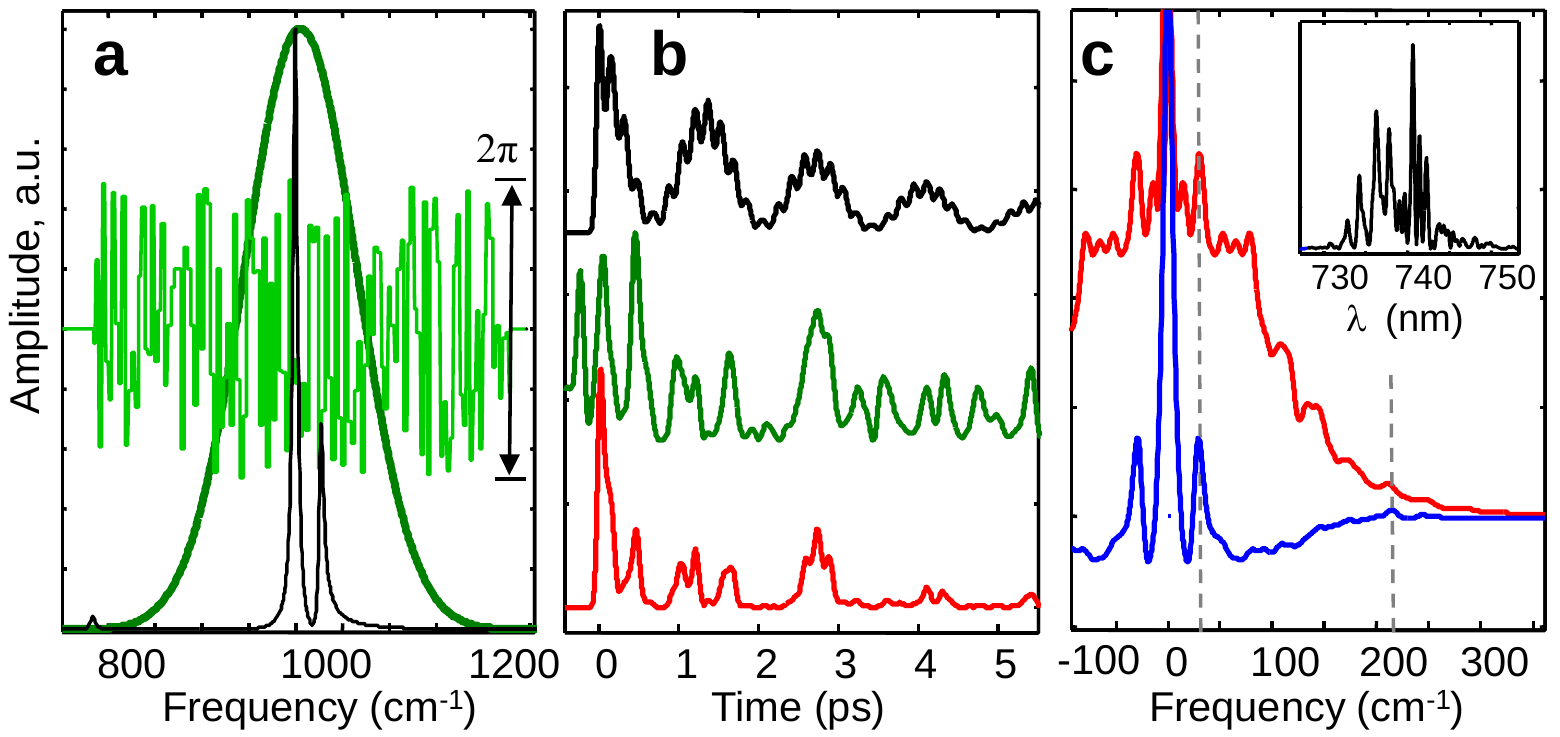}
\caption{
Numerical simulations of the noise auto-correlation spectroscopy with CARS. \textbf{a}, The frequency bandwidth of the probe pulse, represented by the broad gaussian envelope, is much broader than the width of and the distance between the vibrational resonances of Toluene (black curve). White noise is applied to the spectral phase of the probe pulse via random phase jumps of 0-to-2$\pi$ radians every 1 cm$^{-1}$. This frequency step determines the resolution of the NASCARS technique. \textbf{b}, Time domain representation of the decaying molecular vibrations (upper black), incoherent probe pulse train obtained by applying the spectral phase noise to the broadband probe (middle green), and the resulting anti-Stokes pulse train (lower red). Being a product of the vibrational amplitude and the probe field amplitude, the output anti-Stokes field acquires spectral correlations due to the periodic vibrational modulation superimposed onto the uncorrelated input noise. \textbf{c}, Auto-correlation of the calculated noisy anti-Stokes spectrum (shown in the inset, and exhibiting no clear evidence of the Raman modes) with a single realization of the phase noise (upper red) and after averaging over 100 noise realizations (lower blue). The strong peak at 27 cm$^{-1}$ and weak peak at 218 cm$^{-1}$, marked by the dashed lines, correspond to the beating of the vibrational modes of Toluene at 782, 1000 and 1027 cm$^{-1}$.}
\label{Fig2}
\end{figure}

The required amplitude noise in the spectrum of $\mathcal{E}_{0}$ can be introduced by applying either amplitude or phase noise to the input probe field $E_{0}$. Numerical simulation of the NASCARS measurement with a phase-randomized probe field is shown in Fig.~\ref{Fig2}. In the time domain, the spectral noise breaks the initial short probe pulse into a long and random sequence of smaller pulses - an ``incoherent pulse train'' (Fig.~\ref{Fig2}b). In the presence of a long-lived vibrational coherence, this probe train generates a train of the corresponding anti-Stokes pulses. The anti-Stokes train is an exact replica of the original random sequence of the probe pulses with an additional periodic amplitude modulation superimposed on it by the vibrational field of the molecule. It is this amplitude modulation that introduces correlations in the otherwise uncorrelated noise in the observed anti-Stokes spectrum, shown in Fig.~\ref{Fig2}c. High spectral resolution is achieved when the probe train duration is comparable with the decay time of the vibrational coherence.

\begin{figure}
\includegraphics[width=1.0\columnwidth]{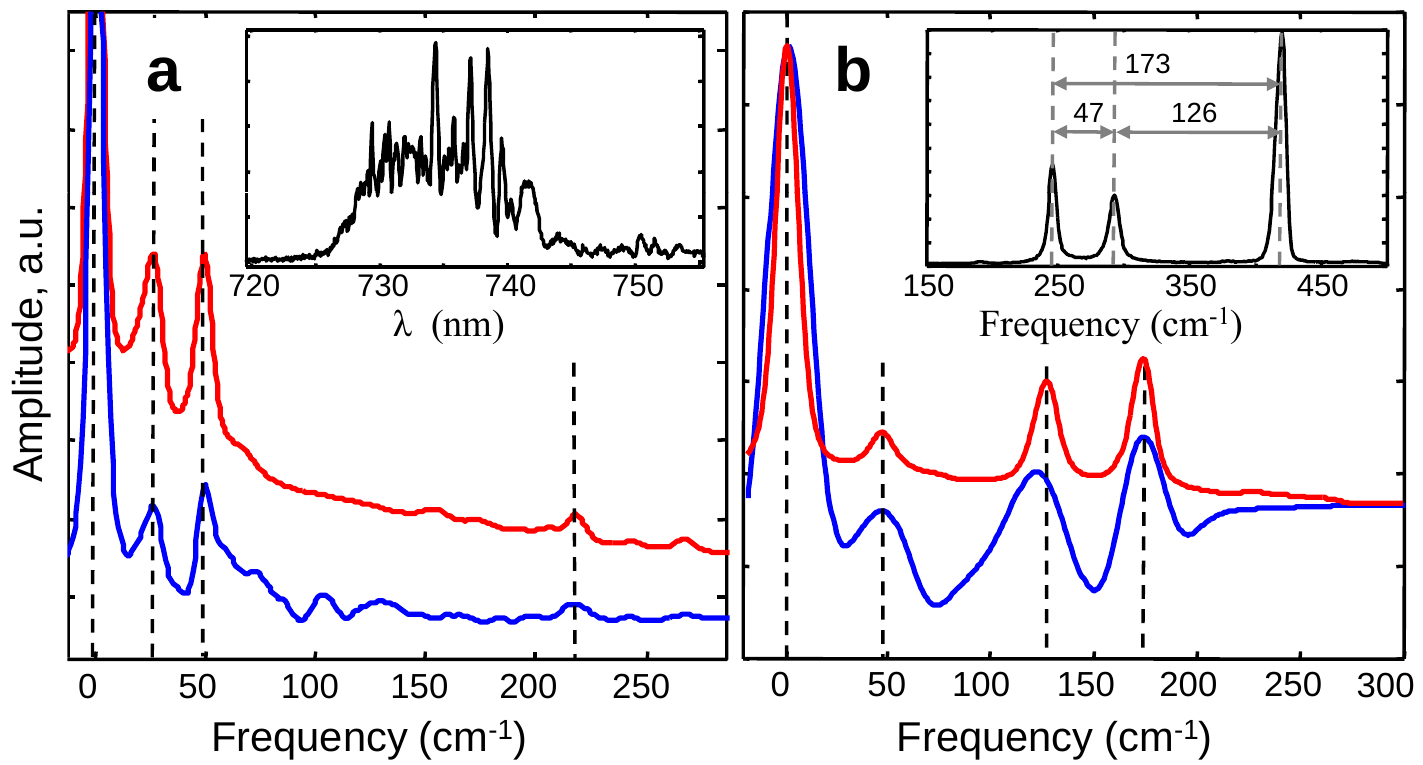}
\caption{
Experimental NASCARS spectra (lower blue) compared with the auto-correlation of the spontaneous Raman spectra obtained by the conventional technique with a separate nanosecond laser system (upper red). \textbf{a}, Mixture of Toluene and ortho-Xylene with the relevant Raman shifts of 782, 1000, 1027, 982 and 1049 cm$^{-1}$. Dashed lines point to the strong beating signals at 27, 49 and 218 cm$^{-1}$. The uncorrelated spectral phase noise was introduced to the probe field through the pulse shaper (Fig.~1b). To suppress the effect of the residual non-resonant background, which contributes to the broad pedestal of the anti-Stokes spectrum (shown in the inset), every two consecutive spectra have been subtracted from each other before an auto-correlation was calculated. The observed linewidth is determined by the convolution of the real resonance width of $\sim$3 cm$^{-1}$ and the spectral resolution of the pulse shaper ($\sim$3.5 cm$^{-1}$) which determines the correlation length of the applied spectral noise. \textbf{b}, Single-pulse NASCARS results for the sample of liquid CBrCl$_{3}$. Here, a single ultrashort pulse (spectral bandwidth of 30nm \textsc{fwhm}) was focused directly in the medium with no additional scanning or shaping optical components inserted in between the laser and the sample. The noise was applied by inserting a thin sheet of a scattering material inside the compressor stage of the laser system, which resulted in a partial de-coherence of the beam. Vibrational beating at 47, 126 and 173 cm$^{-1}$ is clearly identifiable at the expected locations (marked with the dashed lines). The observed peaks are associated with the three Raman modes of CBrCl$_{3}$ at 246, 293 and 419 cm$^{-1}$, shown on the spontaneous Raman spectrum in the inset.}
\label{Fig3}
\end{figure}

We first demonstrate the new method by implementing it in a typical multi-color noncollinear \textsc{boxcars} geometry\cite{Zheltikov00} (see Fig.~\ref{Fig1}b(i)). The setup consisted of a commercial Ti-Sapphire regenerative amplifier providing the probe pulses at a central wavelength of 800nm, and an optical parametric amplifier generating the pump and Stokes pulses at 1111 nm and 1250nm, respectively. All three pulses were focused into a quartz cuvette containing a liquid mixture of Toluene with ortho-Xylene (see \textbf{Methods} for details). The anti-Stokes spectrum was measured with a CCD-based spectrometer, while the noise was introduced by means of a home made programmable spectral pulse shaper. The auto-correlation spectrum presented in Fig.~\ref{Fig3}a was averaged over 100 realizations of noise. Well resolved side-bands at $\pm27$ cm$^{-1}$ and $\pm49$ cm$^{-1}$ in Fig.~\ref{Fig3}a correspond to the beating frequencies between the strongest excited vibrational modes of the mixed sample. In agreement with our theoretical treatment and numerical simulations, the best performance of NASCARS is obtained with a maximum ($2\pi $ peak-to-peak) phase modulation, while the spectral resolution can be controlled through the pulse shaper by varying the distance between the consecutive phase jumps in the pulse spectrum,\ $\delta \omega _{n}$. An easier way of scrambling the probe phase without the shaper is shown in Fig.~\ref{Fig1}b(ii). When coupled to a standard multi-mode fibre, the probe pulse breaks apart into an incoherent pulse train due to the intermodal dispersion, and generates similar auto-correlation spectra.

Numerous applications in vibrational spectroscopy, microscopy and imaging call for a single-pulse technique, recently developed by many groups\cite{Dudovich02,vonVacano06,Lim05,Paskover07} and based on covering the pump, Stokes and probe wavelengths with a broad spectrum of a single ultrashort pulse. Here we demonstrate a single-pulse NASCARS measurement, based on the following mechanism. If the applied noise introduces only partial de-coherence of the original pulse, the remaining coherent part executes an impulsive pump-Stokes excitation, while the incoherent noisy part of the pulse serves as a probe. To NASCARS advantage, such partial de-coherence can be imposed on a laser beam without an elaborate shaping technology, which typically requires delicate optical alignment, calibration, maintenance, and costly optical components. We implemented a single-pulse NASCARS by simply introducing a light scatterer inside the pulse compressor of our laser system (see Fig.~\ref{Fig1}b(iii)) and \textbf{Methods} for details). Placed in the way of a spectrally dispersed beam, the scattering element - a single sheet of a standard lens paper - transmits only some frequency components of the pulse while blocking the others. Standard pulse characterization shows that in accord with the aforementioned requirement, the output laser pulse consists of an ultrashort coherent part and a long noisy tail. Statistical averaging over the realizations of noise is performed by moving the scatterer between the consecutive readouts of the anti-Stokes spectrum. The resulting auto-correlation spectrum of CBrCl$_{3}$, shown in Fig. \ref{Fig3}b, exhibits three clearly resolved peaks at the expected beating frequencies of the molecule, attesting to the feasibility of the proposed method. Remarkably, the whole setup consists of a single slightly modified light source and a spectrometer, yet the achieved resolution of $~20$ cm$^{-1}$ surpasses the spectral bandwidth of our pulses ($>500$ cm$^{-1}$) by more than an order of magnitude.

The demonstrated implementation simplicity of noise auto-correlation spectroscopy with broadband pulses makes it a powerful tool for molecular identification using an auto-correlation of the vibrational spectrum as an excellent fingerprint. It enables quick retrieval of the vibrational beating frequencies with high spectral resolution and no stringent requirements on the quality of the excitation pulse(s). An option to deliver the pulses through the standard high-core-area multi-mode optical fibre, normally prohibited by the intermodal fibre dispersion, should be especially valuable for the applications in nonlinear microscopy. In contrast to multiplex CARS, the pump, Stokes and probe pulses in NASCARS are \emph{all} spectrally broad. As a result, the distribution of the available laser power between these components is independent of the desired spectral resolution and can therefore be adjusted for better performance, e.g. by varying the applied noise level. In principle, other sources of incoherent radiation can be used as long as the frequency correlation length of their noisy spectrum is within the resolving power of the spectrometer.

\section{Methods}
The experiments in the multi-color non-collinear geometry (fig.\ref{Fig1}b(i,ii)) were carried out in a typical CARS setup which consists of a commercial Ti-Sapphire regenerative amplifier (Spitfire, Spectra-Physics, 1 KHz repetition rate, 10 nm spectral bandwidth (\textsc{fwhm}), 2 mJ output pulse power) generating the probe pulses at the central wavelength of 800nm, and an optical parametric amplifier (Topas, Light Conversion) generating the pump and Stokes pulses at 1111 nm and 1250nm, respectively. All three pulses of $\sim$10 $\mu $J energy each were focused by a lens (focal length of 500 mm) into a quartz cuvette (5 mm optical path) with the liquid sample, where they overlapped in time and in space (\textsc{boxcars} phase-matching geometry). The output anti-Stokes beam was collimated with a 150 mm focal-length lens and coupled into the spectrometer (model 2035, McPherson) equipped with a CCD camera (iDus DV400, Andor) and providing a spectral resolution of $<2$ cm$^{-1}$.

In the first configuration (fig.\ref{Fig1}b(i)), we introduced white noise in the probe pulses using the programmable spectral pulse shaper based on a dual-mask 128-pixel liquid-crystal spatial light modulator (SLM-128, CRI). The resolving power of the shaper was $\sim$3.5 cm$^{-1}$ per pixel, which dictated the smallest frequency correlation length of the applied noise, and therefore determined the resolution of the NASCARS signal. From studying the effects of the phase and amplitude noise independently, we conclude that the phase noise results in superior NASCARS performance. In contrast to the amplitude noise, it does not attenuate the available probe pulse energy and results in a much smaller temporal overlap of the probe train with the pump-Stokes pulse pair, reducing the non-resonant background and increasing the signal quality.

We also demonstrated one simple way of phase scrambling by replacing the pulse shaper with a 3 meter-long piece of standard multi-mode fibre of 64 $\mu $m core diameter (Fig.~\ref{Fig1}b(ii)). A probe pulse, coupled into the multiple fibre modes, breaks apart into an incoherent pulse train due to the intermodal dispersion. The large number of available modes and the high sensitivity of the dispersion of each mode to local mechanical stresses in the fibre, enables an easy and efficient way of statistical averaging over the realizations of noise. New realizations of a random probe train are generated by simply moving the fibre between the consecutive data acquisitions. Matching the temporal duration of the output random train to the decay time of the vibrational coherence by choosing an optimal fibre length assures high spectral resolution of the NASCARS measurement.

For the single-pulse experiments we used ultrashort pulses of higher bandwidth from a different femtosecond Ti:Sapphire regenerative amplifier system (Spitfire Pro, Spectra-Physics, 1 KHz repetition rate, 35 nm spectral bandwidth (\textsc{fwhm}), 3 mJ output pulse power). The wavelength of the amplified pulses was centered at 805 nm. The beam was apertured down to a diameter of 0.5 mm before entering the compressor stage of Spitfire Pro. A thin sheet of a light scattering material (optical cleaning tissue, Thorlabs) was fixed on a rigid frame and mounted on a one-dimensional translation stage inside the compressor (fig.\ref{Fig1}b(iii)). The scatterer was placed after the second reflection from the diffraction grating, i.e. at the location where all spectral components of the pulse are parallel and spatially separated. Different noise realizations in this single-pulse configuration were achieved by moving the scattering sheet perpendicular to the dispersed beam. The energy of the output beam was 10 $\mu $J. The laser beam was focused into a quartz cuvette with a liquid sample (5 mm optical path) by a spherical mirror with 300 mm focal length. The transmitted laser beam was collimated with a lens (150 mm focal length) and coupled into the spectrometer (model 2035, McPherson) equipped with a CCD camera (iDus DV400, Andor). The resolving power of the spectrometer in this case was 1.3 cm$^{-1}$. Two frequency filters were employed for enhancing the sensitivity of the detection. A long-wave pass filter (LP02-785RU-25, Semrock) was used to block the high frequency components of the beam in front of the sample, while a short wave pass filter (SP01-785RU-25, Semrock) in front of the spectrometer prevented it from being saturated by the strong incident pump pulse.

In all our experiments, each measured anti-Stokes spectrum corresponded to a different noise realization. After normalization, each two consecutive spectra were subtracted from one another and an auto-correlation of the differential spectrum represented the result of a single NASCARS measurement. The statistical averaging of the auto-correlation was performed over 100 noise realizations.


\begin{acknowledgements}
The authors are grateful to M. Shapiro and E. Shapiro for helpful discussions. This work has been supported by the CFI, BCKDF and NSERC.
\end{acknowledgements}

\end{document}